# A Theory of Information overload applied to perfectly efficient financial markets[1]

Giuseppe Pernagallo† and Benedetto Torrisi‡

*University of Catania, Department of Economics and Business, 55, 95129, Catania*
*†email: giuseppepernagallo@yahoo.it (corresponding author)*
*‡email: btorrisi@unict.it*

## Abstract

Before the massive spread of computer technology, information was far from complex. The development of technology shifted the paradigm: from individuals who faced scarce and costly information to individuals who face massive amounts of information accessible at low costs. Nowadays we are living in the era of "*big data*" and investors deal every day with a huge flow of information. In the spirit of the modern idea that economic agents have limited computational capacity, we propose an original model using information overload to show how "too much" information could cause financial markets to depart from the traditional assumption of informational efficiency. We show that when information tends to infinite, the efficient market hypothesis ceases to be true. This happens also for lower levels of information, when the use of the maximum amount of information is not optimal for investors. The present work can be a stimulus to consider more realistic economic models and it can be further deepened including other realistic features present in financial markets, such as information asymmetry or "noise" in the transmission of information.

---

[1] Please do not quote without permission from the authors.

## 1. Introduction

The traditional theory assumes that individuals use all the relevant information to form their investment decisions and, consequently, prices describe the relevant features of each security. In this paper, we assess the following question: are we sure that individuals will use all the relevant information in their decisional process? The problem was not so evident some decades ago: before the massive spread of computer technology, obtaining information was hard. Several investors must rely on the work of professional operators, who often knew less about finance than they did. In this extremely uncertain scenario, information was a precious good and its cost was not negligible. With the advent of the "internet era" the quantity of available information grew exponentially; digiting on a search engine the words "Apple stock" produces over 1,160,000,000 related contents. Nowadays we do not face anymore a problem of scarce and costly information, but we must deal with massive amounts of data.

The second evolution that interested the phenomenon was the conception of economic agents. The studies of several authors, such as Simon (Simon, 1955, 1956) or Kahneman and Tversky (Kahneman and Tversky, 1979; Tversky and Kahneman, 1986), posed serious doubts over the reliability of the rationality assumption in economic models and, coherently with many studies on the human brain, favoured the hypothesis of a "*bounded*" rationality. We are firmly convinced that reproducing the evident limits of individuals into economic models, although it is a harder task, yields better results.

The novelty of this work consists in showing how information overload combined with big data affects the decision-making process at the individual level and how this can produce informational inefficiency in financial markets. The result is that even an ideal financial market (without cost of trading and noise, where information is freely available and "good") may produce inefficiencies given that agents are limited in their processing routines. Hence, the present work extends conceptually the findings of Grossman and Stiglitz (1980) and shows how, even if information is free, information overload may occur causing a departure from the efficiency hypothesis. Since in our model we assume that all the available information is freely disposable for all the traders, we tackled the strong form of efficiency, proving that this form is unsustainable when information overload is considered.

The paper is organized as follows. Section 2 describes the concept of efficient market hypothesis, the rational expectation hypothesis and the critique of Grossman and Stiglitz (1980). Section 3 presents the theoretical framework: we enlighten the limits of the rational expectations paradigm and discuss the importance of information overload for realistic models. Section 4 presents the model and its results. Section 5 discusses the possible, present and future, implications of the model. The final section concludes.

## 2. Are informationally efficient financial markets possible?

Informational efficiency is the situation that characterizes a market where prices fully reflect all available information. The Nasdaq glossary defines informational efficiency, or equivalently the "efficient market hypothesis" (EMH), as: "*The degree to which market prices correctly and quickly reflect information and thus the true value of an underlying asset*". We must emphasize that the EMH is not properly a model but a working hypothesis, our purpose is to explain why it may not work when information overload is considered. Fama (1970) wrote that the three conditions that ensure market efficiency are:

(i)     trading securities has no cost of transaction,

(ii)    available information is costless for every agent,

(iii)   there is agreement on the implications of current information for the current price and distributions of future prices of each security.

These conditions are only sufficient and not necessary; for example, if (ii) is violated it is sufficient that a number of investors can access to available information to ensure informational efficiency. The violation of (iii) does not necessarily imply market inefficiency, except the case of investors who can systematically evaluate available information better than investor who use only the information reflected in prices.

This hypothesis is often identified with the assumption of agents that obey to Muthian rational expectations (Muth, 1961), in the sense that "*expectations will be identical to optimal forecast (the best guess of the future) using all available information*" (Mishkin, 2004, p. 147), which means that forecasts can be wrong, but they will be the best ones using all disposable information. Following Mishkin (2004), given the variable X, the theory suggests

$$E(X) = X^{of} \qquad (1)$$

that is the expected value of X is equal to the optimal forecast implementing all the available information ($X^{of}$). If markets are efficient, all "unexploited profit opportunities" will be eliminated (Mishkin, 2004) by investors often named "smart money".

Grossman and Stiglitz (1980, p. 404) showed the flaw in this argument with a simple, but effective, consideration: the EMH states that prices reflect, at any time, all disposable information, but if this is true, informed traders could not use information to earn a return, therefore, why do they spend time and money in gathering information? With their model, Grossman and Stiglitz tried to explain that costless information is not a sufficient condition for informationally efficient markets, it is a necessary condition. If information is simply available for everyone and fully reflected in prices but has a cost, it is useless to make returns on it; traders will not find convenient to acquire information and they will start to trade uninformed, but this behaviour will lead to non-informative prices and the EMH will not hold anymore.

## 3. Theoretical framework

### 3.1. From "Olympic" rationality to bounded rationality

The ultimate end of economic agents is always the satisfaction of human desires, which economically translates into the maximization of a utility function.

Muth (1961) criticized previous approaches, noting that the various expectations formulas proposed were not coherent with how the economy really works and suggested the *rational expectations paradigm*, that is "*an approach that assumes that people optimally use all available information—including information about current and prospective policies—to forecast the future*" (Mankiw, 2009, p. 582). This implies that agents may be wrong but over any long period, learning from past mistakes, they will be right, on average. The rational expectation hypothesis was further deepened by Lucas (1972, 1976) and Sargent and Wallace (1975, 1976). The famous Lucas critique (Lucas, 1976) shows that economic policies cannot pursue the estimated targets because once they are implemented by the policy maker, agents will change their behaviour. It is possible to deceive economic agents in order to get the desired result, but as Abraham Lincoln said, "*you can fool all of the people some of the time, you can fool some of the people all of the time, but you can't fool all of the people all of the time*" (Friedman, 1975).

In 1979, Kahneman and Tversky proposed their *prospect theory*, showing several flaws in the traditional expected utility theory. Choices among risky prospects produce various effects that are inconsistent with the utility theory. For example, the *certainty effect* indicates the fact that individuals overvalue outcomes that are obtained with certainty in comparison with merely probable outcomes. The work of Kahneman and Tversky was an important step towards a new approach to economics: in a world where agents are not perfectly rational, the extreme assumption of a sort of "*Olympic rationality*" (Fagiolo and Roventini, 2012) becomes inadequate. Tversky and Kahneman (1986) also criticized the idea that learning from errors will ensure that irrational decision makers will be driven out by rational ones; albeit learning and selection tend to improve efficiency, "*no magic is involved*", and the strict conditions required by effective learning do not resemblance to how agents actually behave.

Simon (1955) discussed the importance of reconsidering the idea of the "economic man", conceived by the traditional theory as completely aware of the relevant aspects of his environment (*global rationality*), in the light of the limited access to information and the true computational capacities that characterize agents. Simon (1956, 1991) defined a new paradigm, called *bounded* *rationality*: decision makers often possess scarce information, they face evident computational limits and have limited time to decide; therefore, economic agents "*adapt well enough to "satisfice"; they do not, in general, "optimize""* (Simon, 1956, p. 129) and, consequently, they often use rules of thumb.

One of the first models that proceeded in a different way from traditional economics was proposed by Arthur (1994). In the "*El Farol Bar problem"*, agents decide whether to go to the bar every Thursday night. If too much agents go to the bar, it becomes too crowded and chaotic. An agent has to decide if it is better to stay at home or go to the bar, conjecturing simultaneously what will be the choice of the other agents. Arthur presents a problem of bounded rationality, where these bounds descent principally from imperfect information and limited computational power.

Keynes already exposed the "not-too-rational" nature of investors in his "*General Theory of Employment, Interest, and Money*" (1936) suggesting that professional investors play a guessing game, similar to a common newspaper game (the so-called *beauty contest*) in which competitors have to guess the six prettiest faces from several photographs, choosing the one considered the more likely to be chosen by the other participants.

The reason of the success of traditional models relies on the fact that building models of rational and unemotional agents is easier than building models of "*quasi-rational emotional humans*" (Thaler, 2000); furthermore, a model with rational agents is more manageable than assuming a world populated by heterogenous agents. But as Thaler predicted, we are witnessing an evolution: from *Homo Economicus* to *Homo Sapiens*.

## 3.2. Information overload: a realistic feature for a realistic model

Information gained a huge role in economics over years. The works of Akerlof (1970), Spence (1973) and Rothschild and Stiglitz (1976), among the vast literature on the issue, finally gave a prominent role to information in economics. However, in these works huge emphasis is given to the nature of information per se, and very little is said about how agents gather this information and successfully use it.

In the era of "*big data*", investors deal every day with a huge flow of information, therefore, an economic model must provide an explanation on how agents process this massive amount of data. The term "information overload" was popularized by Alvin Toffler in 1970 in his classic book "*Future Shock*". Among the various definitions, one suitable for economic and financial studies is that information overload occurs when "*the amount of input into a system exceeds the processing capacity of that system*" (Milord and Perry, 1977, p. 131). The causes of this input saturation arise from the presence of too many inputs for the system to cope with, or because two inputs, say A and B, are presented successively such that both the inputs cannot be adequately processed. The system must set priorities to adapt and decide if it is the case to process A first and keep B in abeyance or sacrifice one input for the other one (Milgram, 1970). This descents from the fact that decision makers have limited processing capacity.

Several studies on the human brain provided evidences on our limited information elaboration capacity. Miller (1956) argued that every individual possesses a finite span of immediate memory estimated to be, for several reasons, about seven items in length. These considerations are the basis for the famous "*magical number seven, plus or minus two*" theory, which states that there is an upper limit of seven plus or minus two elements on our capacity to elaborate information. In a famous experiment carried out by Kaufman et al. (1949), random patterns of dots were flashed on a screen for 1/5 of a second and the individual's task was to count how many dots there were. On patterns containing up to five or six dots the subjects did not make errors, whereas the performance with more dots was very different. Below seven the subjects were said to "*subitize*"; above seven they were said to "*estimate*", that is what is called by Miller the "*span of attention*". What emerges from these studies is that individuals possess limitations (computational power, time, resources etc.) whereas information does not. Klingberg (2000, p. 95) wrote that "*we can only retain a limited amount of information in working memory (WM), and when we try to perform several tasks at the same time, performance deteriorates*".

The model proposed in this paper is based on these considerations, but it is also deeply inspired by the original work of Shannon (1948), who posed as the fundamental problem of communication, the reproduction, either exactly or approximately, of a message from one point to another. He conceived communication systems as perturbed by noise, furthermore, any communication system presents a maximum rate of information that can be carried over a channel, known as *channel capacity*[2]. The "*cognitive cone*" (Figure 1) describes the process of information acquisition using these elements.

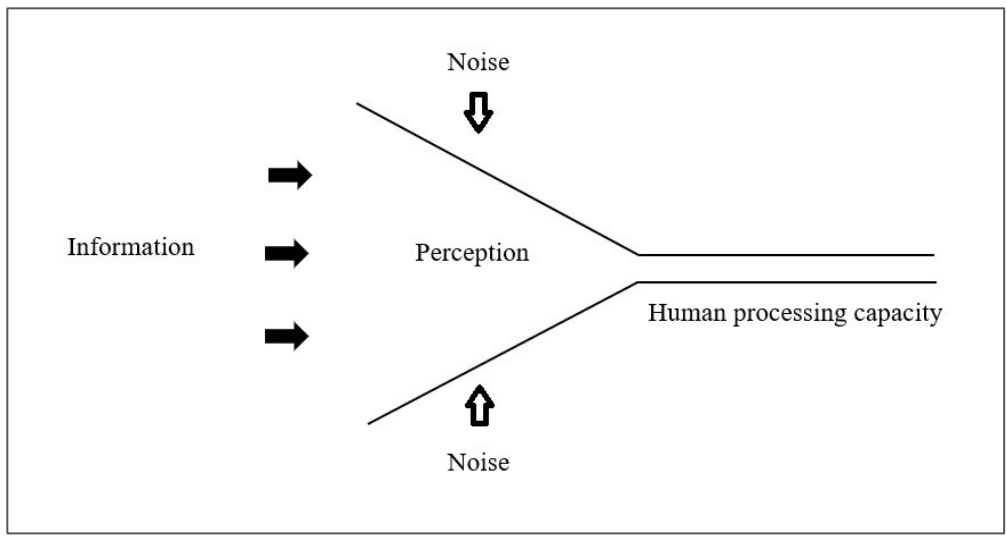

**Fig. 1.** The "cognitive cone".
Source: our elaboration.

The idea is that an agent can perceive more information than he can process. Perception is represented by a cone whereas the shrinkage is the limited processing capacity of human organism (which act as the channel capacity); as in Shannon's model there is a source of noise. The representation of perception as a bigger entity than human processing capacity needs further explanation. We can consider an example (inspired by Kaufman et al., 1949), and the reader can take part to this test. Figure 2-(a), shows a square containing several dots, the task is to try to count the exact number of dots looking at the picture only three seconds. Someone will be able to guess the exact number, others will fail, but everyone can perceive the picture in its entirety. We can use the same rationale to show how information overload reduces our elaboration capacity and makes harder the task. Figure 2-(b) works in the exact same way; with more dots and a more chaotic pattern we are still able to visualize correctly the entire picture, but now our guess is more the result of luck than of a counting.

The argument seems to be obvious, but it is extremely important and less emphasized, especially in financial markets, where a massive amount of information is transmitted almost instantaneously (e.g. intra-day data). Too much information can also lead us to make wrong considerations. Using the same example (and a little bit of philosophy), if we fulfil the square entirely with black dots, and present the figure to an unaware individual, he will say that the image is a black square when, in reality, it is a square containing a certain, huge, number of black dots.

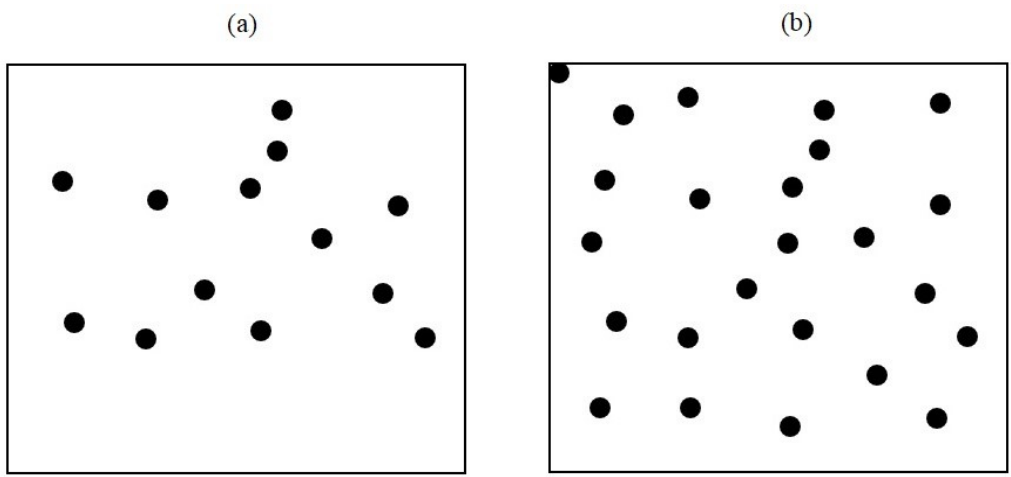

**Fig. 2.** Visual tests to show the discrepancy between perception and processing capacity (a) and how information overload reduces our processing capacity (b).
Source: our elaboration.

## 4. The model

### 4.1. Setup and assumptions

The model here proposed follows a more logical argumentation rather than a mathematical one to show the importance of information overload when assessing the EMH.

**Definition 1.** *In an efficient market prices always reflect all the available information, this descents from the agents' rational expectations assumption. Therefore, a market will be considered efficient if: (a) all the agents use all the available information, or (b) enough traders act fully informed.*

For our purposes, in case (b) we do not need to quantify the number of operators acting fully informed.

**Definition 2.** *Muthian or rational expectations hypothesis assumes that agents optimally use all available information (Mankiw, 2009).*

Definition 2 mathematically translates into the following expressions

$$R_s = R^{of} + \varepsilon \qquad (2)$$

$$E[R_s] = R^{of} \qquad (3)$$

with $R_s$ the return on security $s$, $R^{of}$ the rational expectation on $R_s$ and $\varepsilon$ a white noise process independent from the information possessed. (3) descents immediately from (2) taking the expectations to both sides.

Individuals will gather information only if they can use it to take better positions in the market than the positions of uninformed traders (Grossman and Stiglitz, 1980). We want to show that, even if the traditional assumptions hold, when information overload is considered, informational inefficiencies arise. Our purpose is to prove the following three conjectures.

**Conjecture 1.** *When agents are endowed with Muthian expectations and information is costless, the EMH holds.*

**Conjecture 2 (extension of the Grossman-Stiglitz argument).** *Even if information is costless, when information overload occurs, markets may depart from the assumption of efficiency.*

**Conjecture 3.** *Even if information is costless, when information becomes infinite and overload occurs, markets are not informationally efficient.*

In all the cases, we consider a market where all information is completely free and accessible (to avoid the Grossman-Stiglitz paradox), i.e. condition (ii) holds (see section 2.) and there are no transaction costs (condition (i) holds). Assumption (ii) is not so unrealistic: nowadays information is almost completely accessible via internet, and we can also assume that the cost of buying a computer and of accessing to an internet connection are neglectable. The insider trading normative avoids (almost in theory) the existence of private information spendable to earn excess returns. Finally, there is unanimous agreement on the implications of current information for the current price and distributions of future prices of any security (condition (iii) holds).

Hypothesize that in this market an investment in a security $s$ can yield a positive amount of dollars, $W_s$, with probability $\lambda$ (*probability of success*), and a loss, $L_s$, with probability $(1-\lambda)$, such that the expected return from this investment is

$$E[R_s] = \lambda \, W_s - [1 - \lambda] \, L_s. \tag{4}$$

If the EMH holds, $\lambda$ should not depend from the information possessed because nobody should be able to use the available information to earn excess returns; anyway, in order for investors to be fully informed, information should be "useful" even if it is free. To avoid this impasse, we can think of $\lambda$ as a subjective probability: investors will perceive the available information, $i$, as something useful to increase their probability of success, in particular we can model $\lambda(i)$ as a function with first derivative positive and second derivative negative, or $\lambda'(i) > 0$ and $\lambda''(i) < 0$. The idea is that the probability of success is increasing at decreasing rate because the first amount of information provides to the investor the higher perception of usefulness. Information, $i$, is a continuous variable with values between 0 and the maximum amount of information available in the market, $i_{max}$. For $i = 0$ an investor thinks to have no success because a minimal level of information is required to invest (for example, the knowledge of a trading platform). Assume also that all the available information is "relevant" or there is no "bad" information and that agents are risk neutral. We neglect, for simplicity, the role of noise in the transmission of information. The model is made by two periods: in $t = 0$ the agents choose simultaneously the level of information and in $t = 1$ we can observe two possible outcomes: the market is efficient, or the market is not efficient. Finally, in this model we neglect the role of electronically-based trading. With these assumptions the model produces consistent results even in the presence of electronically-based trading. Indeed, in the first scenario, when agents have infinite elaboration capacities, the addition of electronically-based trading is useless; in the scenario with limited elaboration capacities, electronically-based trading can help to overcome information overload but when information becomes infinite, it results, again, useless because computers cannot store or elaborate infinite information.

## 4.2. Scenario with Muthian expectations

Assume that agents are endowed with Muthian expectations; the expected utility for a trader that invests in a security (we omit for simplicity the subscript $s$) can be written as

$$\text{E}[\text{U}(i)] = \lambda(i)\, \text{W} - [1 - \lambda(i)]\, \text{L}. \tag{5}$$

The maximization problem and its solution for a trader are

$$\max_i \text{E}[\text{U}(i)] \Rightarrow i = i_{max} \Rightarrow \text{E}[\text{R}] = \text{R}^{\text{of}}. \tag{6}$$

In this scenario every trader will choose a level of information equal to $i_{max}$, the maximum amount of information available, and will face $\lambda(i_{max})$. It is important to note that $\text{E}[\text{U}(i)]$ and $\text{E}[\text{R}]$ may differ because while $\text{E}[\text{R}]$ can be considered as how the "world" is, based on objective probabilities, $\text{E}[\text{U}(i)]$ is how investors perceived the "world", based on subjective probabilities. However, since traders take their investment decision considering all the available information $\text{E}[\text{R}] = \text{R}^{\text{of}}$, where this equality follows tautologously from Definition 2. Note that in $t = 0$ we cannot say nothing about $\text{R}^{\text{of}}$: in this stage agents only evaluate how much information they need based on $\text{E}[\text{U}(i)]$. In $t = 1$ traders form their best guess and since they act fully informed, prices always reflect all the available information: the EMH is unavoidably true. This proves Conjecture 1.

## 4.3. Scenario with information overload

We introduce another function, $\xi(i)$, which models the capacity of a trader to elaborate information. This function is based on the "*cognitive cone*" (section 3.2.), hence, we assume that $\xi'(i) > 0$ and $\xi''(i) > 0$, i.e. the elaboration cost function is increasing at increasing rate; realistically, it should be different for each trader: this introduces a source of heterogeneity among agents. This function catches the fact that cognition as the act of memorizing, thinking, searching or obtaining the expertise of other agents is certainly costly (Tirole, 2015). The elaboration cost function is zero for $i = 0$, which is reasonable because if a trader has no information to elaborate, she will not suffer from elaboration cost. We do not need for our purposes to explicit the probability of success and the elaboration cost function. The new expected utility function (7) differs from (5) and must consider this new unavoidable cost

$$\mathrm{E}[\mathrm{U}(i)] = \lambda(i)\,\mathrm{W} - [1 - \lambda(i)]\,\mathrm{L} - \xi(i). \qquad (7)$$

With the new expected utility function (7), we are no longer sure that a trader will always use all the available information because the elaboration cost can cause the expected utility function to decrease before $i_{max}$. Figure 3 illustrates the argument: the elaboration cost causes the expected utility function to decrease before $i_{max}$; this trader will not use all the available information because her optimal level of information is $i^* < i_{max}$. This idea is similar to the processing cost exposed in Persson (2018). In that context the decision maker decides whether to process cues based on the cost $q_R > 0$, however, as she receives more cues, processes fewer, because her expected utility first increases and then decreases, as illustrated in this model.

This situation is not totally destructive for the EMH, indeed, if for a sufficient number of investors $i_{max}$ is lower than $i^*$, the EMH holds because these investors will act totally informed to maximize their expected utility function. This means that traders able to process information at low costs can ensure informational efficiency. This proves Conjecture 2.

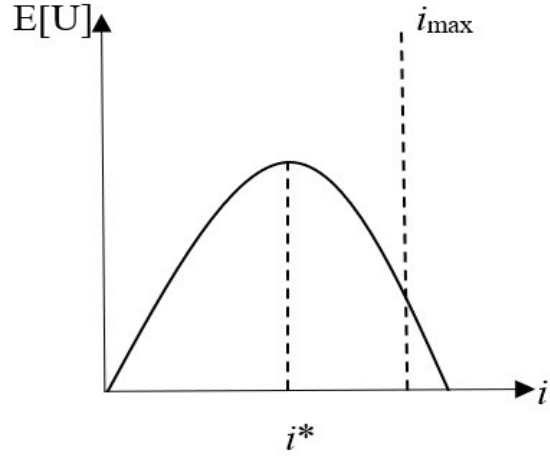

**Fig. 3**. A possible expected utility of a trader affected by information overload.

Source: our elaboration.

Finally, imagine an extreme scenario where the amount of information is infinitely big, or $i \rightarrow +\infty$, therefore

$$i \rightarrow +\infty \Rightarrow \xi(i) \rightarrow +\infty \qquad (8)$$

$$i \rightarrow +\infty \Rightarrow \lambda(i) \rightarrow 1 \qquad (9)$$

it is obvious that when the information amount is infinite, the cost of elaboration is also infinite, whereas the probability of success assumes its maximum value of 1. In this case it results that

$$i \rightarrow +\infty \Rightarrow \mathrm{E}[\mathrm{U}(i)] \rightarrow -\infty. \qquad (10)$$

In this scenario nobody will use all the available information -but it can happen also for lower levels of information- with the consequence that the EMH ceases to be true: if nobody acts fully informed, prices cannot reflect all the available information. This proves Conjecture 3.

## 5. Discussion

With the presented model we have answered our initial question: are we sure that individuals will use all the relevant information in their decisional process? The answer is clearly no. A first conclusion of the model is that informational efficiency, in presence of massive amounts of data (non-redundant, qualitatively good and without noise), depends strictly on the nature of investors. If enough investors can elaborate information at low costs, the EMH may hold. It must be noted that this model is more than a model of financial illiteracy. Albeit educated individuals will exhibit better processing capacities, they remain humans with a maximum capacity of elaboration. The model also shows that when the amount of information becomes too big to be efficiently processed, the EMH ceases to be true, even if all the classical (and ideal) conditions of the model are respected.

Although the model can describe different situations of decision-making, it is thought and applied to financial markets because they are the perfect environment for information overload. It is almost impossible to find another market where information spreads in milliseconds and where decision-making must happen at fast rates. In financial markets more than in other markets the issue of efficiency and the consequences of information overload are indissolubly linked.

Which of the two scenarios is more plausible (investors able to process all the information or investors that suffer from overload) is unsure. On one hand, one can argue that nowadays, with high-frequency trading and the growing number of specialized traders, the elaboration of data may suffer little of information overload. The use of algorithmic procedure consents to elaborate information in fractions of a second (650 microseconds) as they scan large datasets looking for the better opportunities (Spira, 2011). Beddington et al. (2012) think that the development of computer-based trading makes market prices more efficient, improves liquidity, and reduces transaction costs, thus making more reliable the hypothesis of an efficient market. On the other hand, someone may argue that it is practically impossible to fully handle all the information. For example, Blasco and Corredor (2016) found out that investors (informed and uninformed) react to new information provided by financial analysts probably to avoid extra costs related to a long decision-making process. Saxena and Lamest (2018) noted that the amount of available data to managers often exceeds their ability to effectively use these data, with the consequence that part of the information remains underused. We can surely assert that information overload affects a large part of investors, at least, those who are not sufficiently skilled. When investors face information overload, they tend to substitute complex strategies with heuristics. "*A heuristic is a strategy that ignores part of the information, with the goal of making decisions more quickly, frugally, and/or accurately than more complex methods*" (Gigerenzer and Gaissmaier, 2011); the problem with this kind of reasoning is that, if all traders act in this way, informationally efficient markets are not possible, and investors can take bad investment decisions.

There are very few studies that consider the role of the phenomenon in finance; an important investigation was carried out by Agnew and Szykman (2005). The authors found that individuals with a low level of financial knowledge suffer from overload, which leads them to take the path of least resistance, the default option in defined contribution retirement plans, making very often the wrong investment decision. When decision makers are unable to make a mindful investment choice, they opt for the easiest alternative. It is exactly what happens when a customer can choose among a great variety of pizzas from a menu: she can spend a lot of time in examining every possibility, take some "risk" trying a new pizza, or simply order a Margherita. Agnew and Szykman indicated three sources of overload in addition to a poor financial knowledge: how information about alternatives is presented to the investors, the number of investment choices offered in a pension plan, and the similarity among the funds offered, because similar funds are harder to differentiate from each other.

Spira (2011) pointed out the relevant consequence of information overload in explaining financial markets crises. The recent financial crisis derived from a financial industry that became too complicated to calculate adequately the risks of an investment. It is the case of the so-called "derivatives on steroids", which are derivatives on derivatives. The more complex the derivative, the less transparent its risk evaluation process.

The inadequacy of the old notion of efficiency shows that it must be updated, maybe considering financial markets as governed by biological laws, where competition or interactions among the agents resemble the game of survival that biological organisms face (Lo, 2017). Conceiving agents as biological organisms and not as computers entails a better representation on how markets really work.

The model here exposed can be made mathematically more complex and realistic considering other aspects interacting with overload, neglected in this work for simplicity. For example, information asymmetries, sources of noise in the transmission process or misleading information (e.g. fake news) can be contents of future works.

## 6. Conclusions

This paper shows the importance of including more realistic assumptions in economic modelling. Following the modern idea of agents with limited capacities (Simon, 1955, 1956, 1991; Kahneman and Tversky, 1979; Tversky and Kahneman, 1986; Arthur, 1994; Thaler, 2000) we have used the phenomenon of information overload to provide new insights into the possible causes of informational inefficiency. The agents that populate the model have limited processing capacity coherently with many studies over the human brain (Kaufman et al., 1949; Miller, 1956). To sum up, our model can explain departures from the traditional assumption of informational efficiency introducing an elaboration cost function. This function, which is increasing at an increasing rate, introduces information overload: this occurs when the cost from elaborating another unit of information is not compensated by the increment of the subjective probability of success. The assumption of individuals that use all the available information is far from realistic, especially when information is "too much". Agents choose the optimum level of information given their probability of success and their elaboration cost function; if the optimum is in correspondence of a lower level of information than the maximum amount available, then a trader will not act fully informed. If information becomes infinite, nobody will act fully informed and the EMH ceases to be true. This conclusion implies that: traders with low elaboration cost are the most likely to act fully informed and to ensure the efficiency; in the modern society of "*big data*" information overload becomes a persistent phenomenon with the consequence that departures from efficiency are most likely to occur; individuals will use all the available information only when they can effectively handle it, this probably explains why very often traders prefer rules of thumb or the default option to complex investment strategies (Agnew and Szykman, 2005).

This framework can be extended in several ways. It is possible to add more realistic features, for example, what if we include also a distinction between "good" and "bad" information? What if we include a source of "*noise*" during the process of transmission? These are only few suggestions for further developments of the model.

Since it first appeared, the efficient market hypothesis has been considered the Gordian Knot of finance. Even if we think that the market is not "invincible", being able to continuously "beat the market" seems to be unlikely. The aim of the work is to criticize the use of the word "perfect" in association with something regarding the human behaviour. Humans are not perfect, and so markets. Our wish is to assist soon to a definitive consolidation of a paradigm where realism takes the place of idealism and where economic agents finally act as humans.

**Funding**

This research did not receive any specific grant from funding agencies in the public, commercial, or not-for-profit sectors.